\documentclass[twoside,english,final,5p,times,twocolumn]{elsarticle}
\usepackage[T1]{fontenc}
\usepackage[latin9]{luainputenc}
\usepackage{babel}
\usepackage{float}
\usepackage{units}
\usepackage{amstext}
\usepackage{graphicx}
\PassOptionsToPackage{version=3}{mhchem}
\usepackage{mhchem}
\usepackage{wasysym}
\usepackage[unicode=true]
 {hyperref}

\makeatletter

\providecommand{\tabularnewline}{\\}

\journal{Ultramicroscopy}

\makeatother

\begin{document}

\begin{frontmatter}{}

\title{Calibration-sample free distortion correction of electron diffraction
patterns using deep learning}

\author[uvic,camtec]{Matthew~R.C.~Fitzpatrick\corref{cor1}}

\ead{mrfitzpa@uvic.ca}

\author[uvic,camtec]{Arthur~M.~Blackburn}

\author[uvic,camtec]{Cristina~Cordoba}

\cortext[cor1]{Corresponding author}

\address[uvic]{Department of Physics and Astronomy, University of Victoria, BC V8W 2Y2, Canada}

\address[camtec]{Centre for Advanced Materials and Related Technologies, University of Victoria, BC V8W 2Y2, Canada}
\begin{abstract}
The accuracy of the information that can be extracted from electron
diffraction patterns is often limited by the presence of optical distortions.
Existing distortion characterization techniques typically require
knowledge of the reciprocal lattice of either the sample of interest
or a separate calibration sample, the latter of which would need to
be swapped in, thus adding time and inconvenience to an experiment.
To overcome this limitation, we develop a deep learning (DL) framework
for measuring and correcting combinations of different types of optical
distortion in CBED patterns. Quantitative performance tests of our
DL model are conducted using testing datasets of artificial distorted
CBED patterns of $\ce{MoS2}$ on amorphous $\ce{C}$, with varying
sizes of CBED disks, that are generated using multislice simulations.
The performance test results of our DL approach are benchmarked against
those obtained using a conventional distortion estimation technique
that uses the radial gradient maximization (RGM) technique and knowledge
of the reciprocal lattice system. While the RGM approach outperforms
our DL approach for the CBED patterns with very small disks, our DL
approach outperforms the RGM approach for the CBED patterns with medium-sized
disks, as well as those with large overlapping disks. The benchmarking
results suggest that our DL approach, which does not require knowledge
of the sample, achieves a good compromise between convenience and
accuracy. We also show how our DL framework can be used to improve
experimental ptychographic reconstructions, and to correct optical
distortion in experimental selected area electron diffraction patterns.
\end{abstract}

\end{frontmatter}{}

\section{Introduction\label{sec:Introduction__1}}

Transmission electron microscopes (TEMs) have played a central role
in acquiring high resolution structural information of crystals in
material science, as well as soft matter materials in microbiology
\citep{Vinothkumar_1,Wei_1,Pennycook_1,Murata_1}. To achieve sub-Ångström
($\unit[<10^{-10}]{m}$) resolution imaging of samples in current
conventional TEMs, one must use generally high-energy electron beams
($\unit[>30]{keV})$ in conjunction with recently developed aberration
correctors \citep{Linck_1}. Furthermore, for beam-sensitive materials,
e.g. biological materials, the transmitted electrons must be recorded
using an advanced high-pixel count direct electron detector with the
greatest possible detection efficiency \citep{Vinothkumar_1}. The
detector advances that have emerged from these initiatives, along
with recent advances in aberration correctors for TEM and cryogenic
sample preparation methods, have made TEMs indispensable tools for
structural investigations related to microbiology, including pathogen
biology, host-pathogen interactions, and drug discovery \citep{Wei_1,Pennycook_1,Murata_1,Linck_1}.

While these technological advancements have enabled high-resolution
imaging in conventional TEMs, they impose financial costs, along with
space and personnel requirements that are prohibitive for many laboratories
\citep{Vinothkumar_1}. By contrast, non-aberration corrected scanning
electron microscopes (SEMs) are smaller, and use electron beam energies
typically below $\unit[30]{keV}$. Recently, Blackburn \emph{et al.}
\citep{Blackburn_1} achieved sub-Ångström imaging of a gold on Amorphous
Carbon (Au/aC) thin film using ptychography with convergent beam electron
diffraction (CBED) data collected by a non-aberration-corrected scanning
electron microscope (SEM) operating in a transmission mode with a
defocused $\unit[20]{keV}$ electron beam. Such microscopes are cheaper
to run and maintain by comparison, and as a result, have become a
popular alternative analtyical tool in engineering, and the micro-
and nano-sciences.

Beyond economic considerations, lower beam energies ($\unit[\le30]{keV})$
should yield higher information per unit damage in TEM and scanning
TEM (STEM) modes, provided that the specimen is sufficiently thin
($\apprle\unit[15]{nm}$) and composed primarily of light elements
\citep{Peet_1}, e.g. carbon. This applies to small proteins in cryo-EM
corresponding with molecular masses below $\unit[100]{kDa}$ \citep{Erickson_1}.
There has been a growing interest in determining the detailed structure
of low-mass ($<\unit[100]{kDa}$) proteins \citep{Shepherd_1}, which
are particularly abundant in nature but difficult to characterize
\citep{Wentinck_1}. Graphene and hexagonal boron nitride, both being
two-dimensional (2D) materials comprising of light atoms, hold great
promise for future electronics due to their structural, chemical,
and electronic properties \citep{Martini_1}. Many heterostructures
comprise of layers of these materials with thicknesses below the $\apprle\unit[15]{nm}$
limit mentioned above \citep{Martini_1}. 

In addition to improved information per unit damage, the typical beam
energies of non-aberration corrected SEMs are well below the threshold
displacement energies (TDEs) of many 2D transition metal dichalgenides
(TMDCs) \citep{Kretschmer_1}. The main bottleneck to the scaling
of the production of devices that integrate 2D TMDCs like $\ce{MoS2}$
is manufacturing readiness \citep{Lemme_1}, which requires imaging
techniques that enable sub-Ångström resolution, at beam energies below
the material's TDE.

While there are many advantages to using non-aberration-corrected
SEMs, typically in such instruments the angular field of view in CBED
mode is largely controlled by a single projector lens that is often
a major source of optical distortions at low beam energies \citep{Blackburn_1}.
Such distortions limit the accuracy of the information that we can
derive from CBED patterns collected in the microscope, including structural
and orientation information, in addition to high-resolution phase-amplitude
imagery via ptychography. Without physical optical elements that can
correct lens distortions, one must then resort to a post data collection,
software-based solution.

CBED patterns are produced using a convergent incident electron beam.
Consequently, if the sample of interest is crystalline, all optical
elements are perfectly aligned in a given CBED experiment, the Ewald
sphere curvature is sufficiently small, the small angle approximation
is valid across the angular field of view of the CBED pattern, and
no optical distortion is present, then a circular diffraction disk
will be centered at every Bragg reflection captured by the recorded
CBED intensity pattern(s). In the presence of optical distortion,
the shapes and centers of these disks (i.e. CBED disks) will be deformed
and displaced respectively. Existing distortion characterization techniques
typically involve estimating the displaced centers of the CBED disks
in a given pattern, using e.g. the radial gradient maximization (RGM)
technique \citep{Muller_1,Mahr_1}, and then performing some kind
of least-squares optimization procedure according to an expected or
assumed reciprocal lattice system. When the reciprocal lattice of
the sample of interest is not known \textemdash{} as is often the
case \textemdash{} one must swap in another sample that has a known
reciprocal lattice, estimate the distortion, and then swap back in
the original sample. This extra step of running a calibration sample
adds time and inconvenience to a CBED experiment. Moreover, accurate
localization of CBED disks using the RGM technique is complicated
by overlapping CBED disks.

To overcome these limitations, we develop a deep learning (DL) framework
for estimating the optical distortion in CBED patterns. Several DL
methods have already been developed for distortion correction problems,
intended for generic use cases, i.e. not specialized for electron
microscopy. Models/frameworks like Radial Distortion Transformer \citep{Wang_1},
Distortion-Aware Representation Learning for Fisheye Image Rectification
\citep{Liao_1}, Distortion Rectification Generative Adversarial Network
\citep{Liao_2}, Simple Framework for Fisheye Image Rectification
\citep{Feng_1}, and GeoNet \citep{Li_1} assume that only one type
of distortion is present in any given input image. Recently, Li \emph{et
al.} \citep{Li_2} developed a two-stage method that combines GeoNet
with a traditional image registration algorithm that is capable of
handling images that contain mixtures of different types of distortion.
However, the method is limited by its dependency on reference images
to perform the image registration. Our DL framework is specially designed
to predict the distortion field of a given CBED pattern according
to the deformities of the CBED disks, rather than the displacements
of the CBED disk centers. Consequently, our approach has the advantage
of not requiring knowledge of the sample of interest. Moreover, our
DL framework is capable of handling generic mixtures of different
types of distortion, and CBED patterns with overlapping disks.

In this paper, we describe in detail our DL framework for distortion
correction of CBED patterns. We present test results quantifying the
accuracy of our DL model and compare it to distortion estimation performed
using the RGM technique. We also present two applications of our DL
framework: First, we correct the distortion in an experimental CBED
pattern (i.e. 4D-STEM) dataset used for ptychography and show that
this preprocessing of the data improves the quality of ptychographic
reconstructions; Secondly, we show how our DL framework can be adapted
for distortion correction in experimental selected area electron diffraction
(SAED).

\section{Methods\label{sec:Methods__1}}

\subsection{Convergent beam electron diffraction\label{subsec:cbed__1}}

Our DL framework is specially designed to correct optical distortion
in CBED patterns. The essential property of CBED is that the sample
is illuminated by a convergent electron beam, which produces diffraction
patterns containing diffraction disks if the sample is crystalline.
Beyond this, CBED experiments vary in terms of the number and types
of optical elements used. Our DL framework is particularly useful
for simple few-lens microscopes that are not aberration-corrected
by hardware. 

\begin{figure}[h]
\noindent \begin{centering}
\includegraphics[scale=0.49]{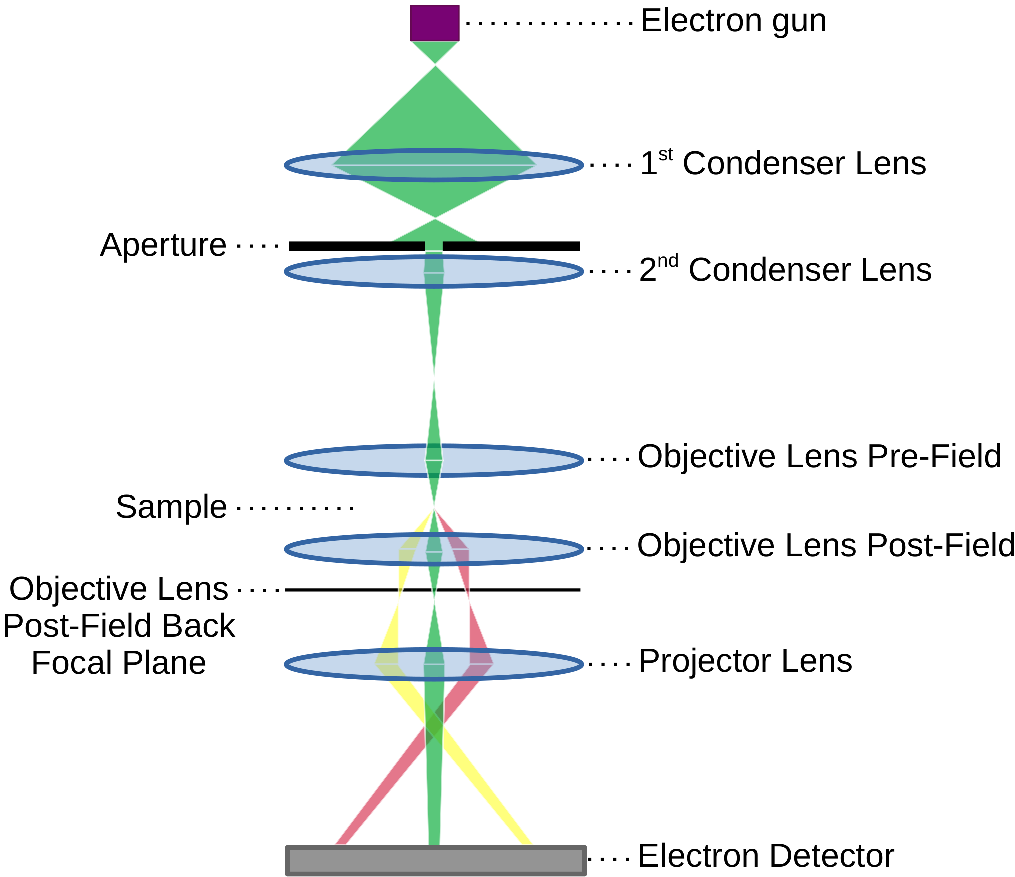}
\par\end{centering}
\caption{A schematic of the modified Hitachi SU9000 optical system in CBED
mode.\label{fig:schematic_of_SU9000__1}}
\end{figure}

In this paper, CBED experiments were performed on a modified Hitachi
SU9000 SEM. Fig.~\ref{fig:schematic_of_SU9000__1} shows a schematic
of the modified Hitachi SU9000 optical system in CBED mode, which
is representative of other simple few-lens non-aberration-corrected
CBED optical systems. This SEM is equipped with a cold field-emission
electron gun, stabilized with a non-evaporative getter pump \citep{Kasuya_1}.
The instrument was modified from a standard design to include a small
projector lens beneath the immersion-objective magnetic lens. This
additional lens improves the accessible range of the effective camera
length, i.e. the magnification in the diffraction plane, which is
realized in combination with varying the objective lens current and
the sample height within the objective lens gap. A hybrid-type pixel
array direct electron detector was added to the SEM to collect the
CBED intensity patterns. This detector, which collects $512\times512$
pixel-count images, is based on the EIGER detector design \citep{Tinti_1},
as provided in a Quadro family camera from Dectris AG, Switzerland.

\subsection{Parameterizing optical distortion\label{subsec:parameterizing_optical_distortion__1}}

Broadly speaking, aberrations in an optical system can be grouped
into two categories: those that produce inhomogeneities in magnification;
and those that affect image resolution. The former are referred to
as optical distortions, whereas the latter include coma and aperture
aberrations like e.g. spherical aberrations. In other words, optical
distortions will not affect the blurriness of an image, but it will
cause deformations in the geometric features depicted by that image.
Therefore, in order to estimate the distortion in a CBED pattern from
the CBED pattern alone, one must analyze some subset of geometric
features in the CBED pattern. Furthermore, in order for a distortion
estimation framework to not require knowledge of the sample of interest,
assumed to be crystalline, one must analyze in particular a subset
of geometric features that does not depend on the sample. This rules
out analyzing the displacement of the centers of the CBED disks in
a CBED pattern as they depend on the reciprocal lattice system of
the sample. In contrast, the shapes of the CBED disks do not depend
on the sample.

If all optical elements are perfectly aligned in a given CBED optical
system, the Ewald sphere curvature is sufficiently small, the small
angle approximation is valid across the angular field of view of the
CBED pattern, and no optical distortion is present, then the shapes
of the CBED disks that are depicted by a recorded CBED intensity pattern
will be near-perfect circles of the same common radius. This is due
to the fact that, under these conditions, the shape of the cross section
of the incident beam at the sample plane, i.e. the ``probe shape'',
is almost perfectly circular. As is suggested by Fig.~\ref{fig:schematic_of_SU9000__1},
for the SEM that we use in this paper, the probe shape is determined
by pre-sample lens distortion effects, and the alignment of the pre-sample
lenses and the second condenser lens (C2) aperture. Typically, the
distortion effects of condenser and objective lenses are negligible
while aperture aberrations and coma are dominant \citep{Hawkes_1}.
Hence, under these circumstances, only misalignment of the aforementioned
optical elements can cause probe shapes that are appreciably non-circular.
For example, an unintended tilt in the C2 aperture can cause elliptical
probe shapes. Unlike pre-sample lenses, projection lenses are typically
distortion dominant while aperture aberrations and coma effects are
negligible \citep{Hawkes_1}. 

The facts laid out in this section thus far form the basis of our
DL framework: assuming that the probe shape is known and that distortion
effects are originating predominantly from the projector lens, we
can estimate the distortion in a CBED pattern from the CBED pattern
alone if we analyze how the shapes of the deformed CBED disks deviate
from the probe shape. For simplicity, we assume that all optical elements
are perfectly aligned, which implies that the probe shape is a near-perfect
circle. 

Our DL model accepts as input a distorted CBED pattern, and returns
as output a representation of the CBED pattern's distortion field.
We describe below how distortion fields are parameterized in this
work. Before doing so, we introduce some convenient notation. To start,
let $E_{\Square}$ denote a CBED experiment of a sample wherein the
optical system is operating at a fixed set of target parameters, and
all of the optical elements used are idealized in the sense that they
do not introduce any optical distortions. Next, let $E_{\wasylozenge}$
denote a CBED experiment that is identical to $E_{\Square}$ except
that the optical elements used introduce a set of optical distortions.
We refer to the set of CBED patterns resulting from the experiment
$E_{\Square}$ as the set of undistorted CBED patterns, and the set
of CBED patterns resulting from the experiment $E_{\Square}$ as the
set of distorted CBED patterns. For simplicity, we describe positions
within CBED patterns using fractional coordinates. First, let $u_{x}$
and $u_{y}$ be the fractional horizontal and vertical coordinates,
respectively, of a point in an undistorted CBED pattern. Secondly,
let $q_{x}$ and $q_{y}$ be the fractional horizontal and vertical
coordinates, respectively, of a point in a distorted CBED pattern.

\begin{figure}[h]
\noindent \begin{centering}
\includegraphics[scale=0.43]{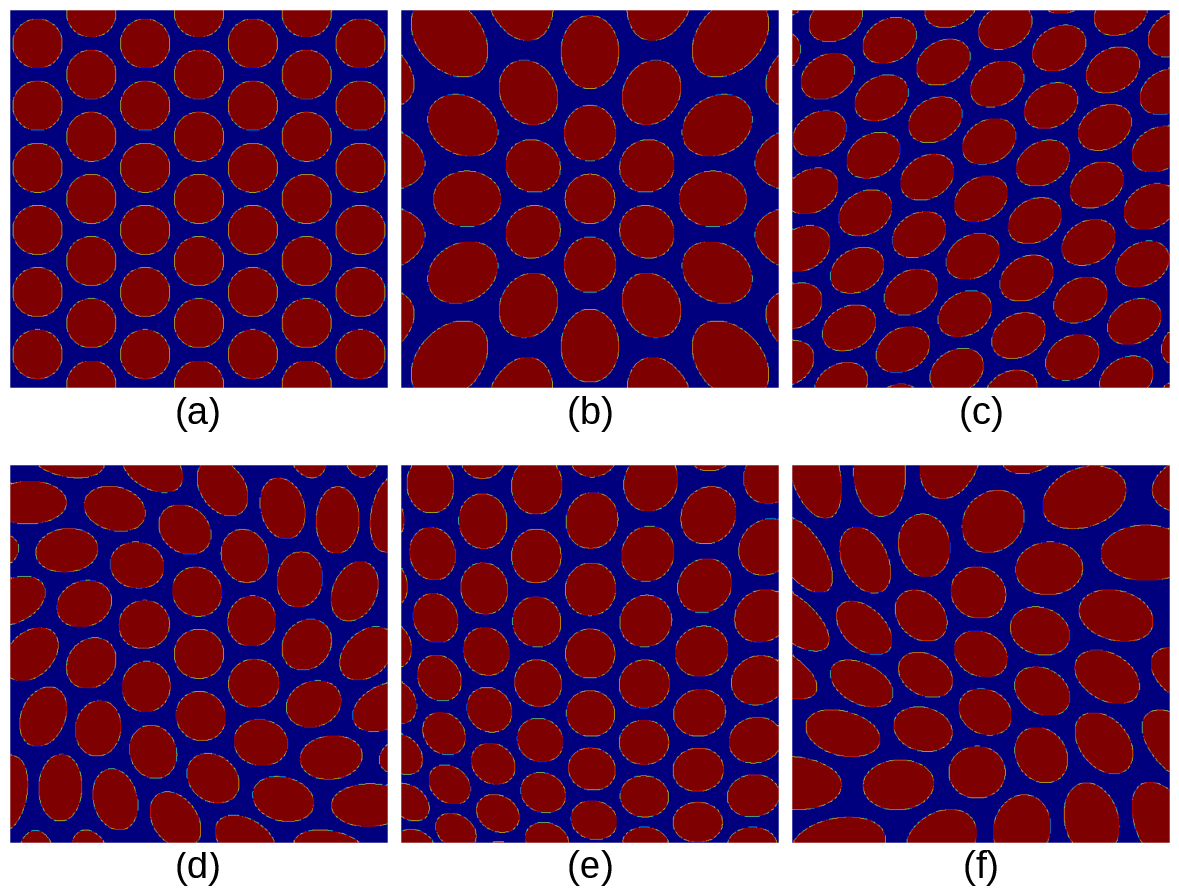}
\par\end{centering}
\caption{Illustrations of the types of distortions considered in our DL framework:
(a) No distortion; (b) Quadratic radial; (c) Elliptical; (d) Spiral;
(e) Parabolic; and (f) Generic mixture.\label{fig:illustrations_of_the_four_distortion_types__1}}
\end{figure}

The distortions introduced by experiment $E_{\wasylozenge}$ can be
described by a coordinate transformation, which maps a given coordinate
pair $\left(u_{x},u_{y}\right)$ to a corresponding coordinate pair
$\left(q_{x},q_{y}\right)$. Let $T_{\wasylozenge;x}\left(u_{x},u_{y}\right)$
be the component of the coordinate transformation that maps $\left(u_{x},u_{y}\right)$
to its corresponding $q_{x}$, and $T_{\wasylozenge;y}\left(u_{x},u_{y}\right)$
be the component of the coordinate transformation that maps $\left(u_{x},u_{y}\right)$
to its corresponding $q_{y}$. Following Ref.~\citep{Brazda_1},
we assume that the coordinate transformation can be parameterized
by a trigonometric series. Complete mathematical descriptions of $T_{\wasylozenge;x}\left(u_{x},u_{y}\right)$
and $T_{\wasylozenge;y}\left(u_{x},u_{y}\right)$ can be found in
the reference guide of $\texttt{DistOptica}$ \citep{Distoptica_1},
a Python library that we have developed for modelling optical distortions. 

While the aforementioned coordinate transformation is expressive enough
to describe a wide variety of different types of distortion, in this
work we restrict ourselves to 4 types of distortion that are prevalent
in diffraction experiments: quadratic radial (i.e. barrel-pincushion),
elliptical, spiral, and parabolic. Fig.~\ref{fig:illustrations_of_the_four_distortion_types__1}
illustrates how these 4 types of distortion alter an image. In $\texttt{DistOptica}$,
these 4 types of distortion are minimally described by a class of
so-called ``standard'' coordinate transformations. Each standard
coordinate transformation comprises of two components which have the
same mathematical forms as $T_{\wasylozenge;x}\left(u_{x},u_{y}\right)$
and $T_{\wasylozenge;y}\left(u_{x},u_{y}\right)$ respectively, but
is constrained such that a given standard coordinate transformation
can be uniquely specified by 8 distortion parameters. These distortion
parameters are: the distortion center $\left(x_{c;D},y_{c;D}\right)$,
the quadratic radial distortion amplitude $A_{r;0,2}$, the elliptical
distortion vector $\left(A_{r;2,0},B_{r;1,0}\right)$, the spiral
distortion amplitude $A_{t;0,2}$, and the parabolic distortion vector
$\left(A_{r;1,1},B_{r;0,1}\right)$. 

As the distortions introduced by an experiment $E_{\wasylozenge}$
can be described by a coordinate transformation, they can be described
equivalently by a distortion field, with components $D_{\wasylozenge;x}\left(u_{x},u_{y}\right)=T_{\wasylozenge;x}\left(u_{x},u_{y}\right)-u_{x}$
and $D_{\wasylozenge;y}\left(u_{x},u_{y}\right)=T_{\wasylozenge;y}\left(u_{x},u_{y}\right)-u_{y}$.
Thus, when we refer to the distortion field of a distorted CBED pattern,
we are referring to $\left(D_{\wasylozenge;x}\left(u_{x},u_{y}\right),D_{\wasylozenge;y}\left(u_{x},u_{y}\right)\right)$.
We refer to distortion fields that can be described equivalently by
``standard'' coordinate transformations as standard distortion fields.

Each data instance in each dataset used to either train, validate,
or test our DL model contains a single artificial distorted CBED pattern,
and the 8 distortion parameters that specify the distortion field
of the artificial CBED pattern, with each distortion parameter being
min-max normalized with respect to some set of data instances. For
a data instance belonging to either the training or validation dataset,
the distortion parameters belonging to that data instance are min-max
normalized with respect to the union of the data instances stored
in the training and validation datasets. For a data instance belonging
to a testing dataset, the distortion parameters belonging to that
data instance are min-maxed normalized with respect to all of the
data instances stored in that testing dataset.

\subsection{Deep learning model\label{subsec:deep_learning_model__1}}

Our DL model accepts as input a distorted CBED pattern, and returns
as output the distortion parameters that specify the estimated distortion
field of the CBED pattern, normalized using the same weights and biases
as those used to normalize the data instances stored in the training
dataset. The model has a encoder architecture that is similar to ResNet
architectures \citep{He_1}, though it has a few notable differences:
First, prior to any convolutions, the CBED pattern is min-max normalized,
then gamma-corrected, then histogram equalized to improve the contrast
of the CBED disks; Secondly, no pooling operations are performed in
our DL model; Lastly, all downsampling operations are performed using
convolutional layers with stride equal to 2. A detailed description
of the architecture of our DL model \textemdash{} which we call ``$\texttt{DistopticaNet}$''
\textemdash{} can be found in the reference guide of $\texttt{EMicroML}$
\citep{Emicroml_1}, a Python library that we have developed that
provides the computational framework for training machine learning
models for tasks in electron microscopy.

\subsection{Training and validation data generation\label{subsec:training_and_validation_data_generation__1}}

A common approach to generating training data for machine learning
problems in electron diffraction is to use a physics-based simulator
to generate artificial diffraction patterns, e.g. the Bloch wave method
\citep{Yuan_1,Kirkland_1} or the multislice method \citep{Munshi_1,Kirkland_1}.
While physics-based simulators can yield training images that accurately
depict experimental CBED patterns, they are typically time consuming
and compute resource intensive. Moreover, in order to generate a diverse
training dataset, one must model tediously a variety of samples of
different lattice types, thicknesses, and defects. There are also
an overwhelming number of optical system model parameters to vary
such as the beam energy, semi-convergence angle, and lens aberrations.

Rather than use physics-based simulators, we use basic mathematical
functions to generate training and validation images that depict the
essential geometric features of CBED patterns: CBED disks are modelled
using circular disk supports, with intra-disk patterns generated using
combinations of plane waves, asymmetric Gaussian and Lorentzian peaks,
and hydrogen-like atomic orbitals; Kikuchi bands are modelled using
basic geometric bands; Background intensities are modelled using combinations
of asymmetric exponential, Gaussian, and Lorentzian peaks; Illumination
supports (i.e. regions outside which electrons are not detected) are
modelled using rectangular, circular, elliptical, and generic blob
shapes; Aperture aberration effects are approximated using Gaussian
filters; Poisson noise is also included. The properties \textemdash{}
including locations, orientations, and length scales \textemdash{}
of the geometric objects used to generate training and validation
images are sampled randomly from various distributions.

\begin{figure}[h]
\noindent \begin{centering}
\includegraphics[scale=0.43]{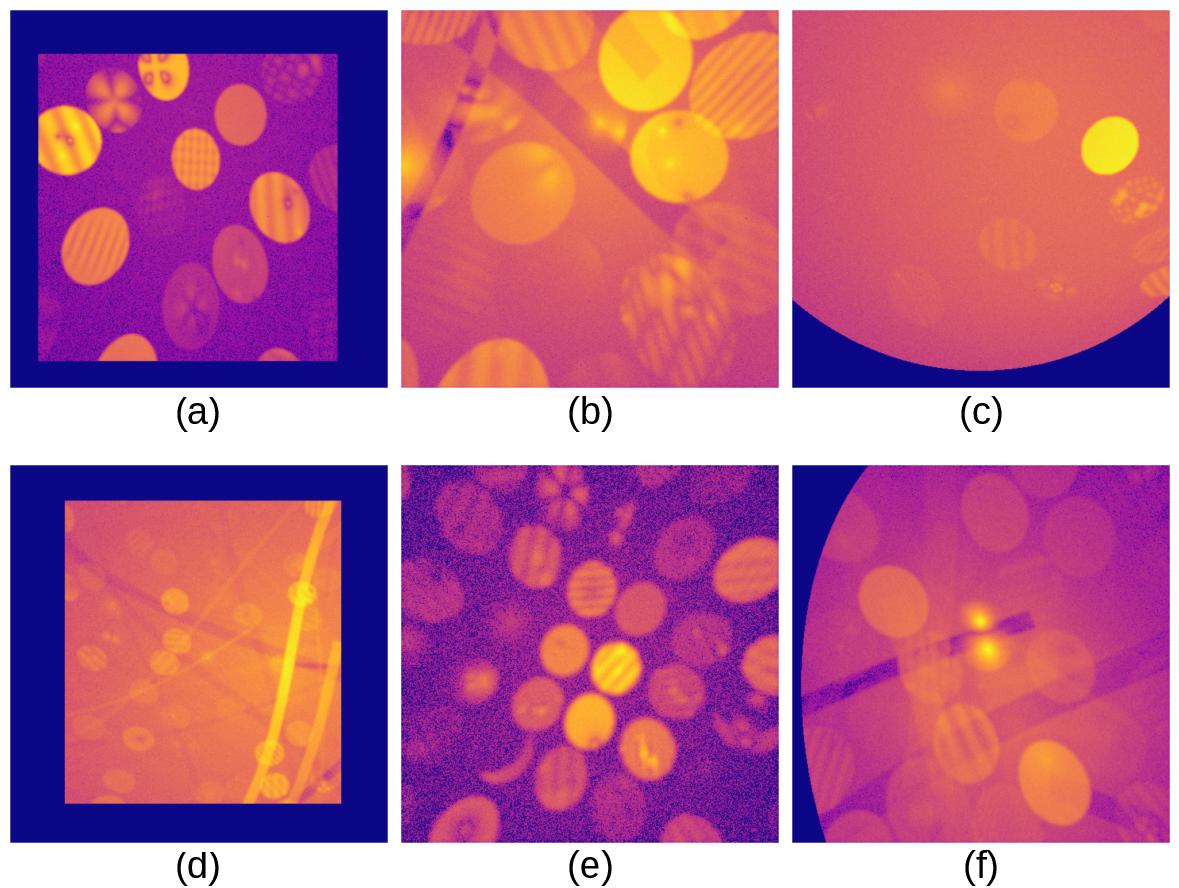}
\par\end{centering}
\caption{Sample of training images, gamma-corrected with a gamma value of $0.2$.\label{fig:sample_of_training_images__1}}
\end{figure}

It is important to stress that our goal is not to generate images
that accurately depict experimental CBED patterns, but to generate
images that depict the essential geometric features of CBED patterns
that can be detected and exploited by a DL model to predict the distortion
field. The most important geometric feature to depict is that the
artificial CBED disks in a given image are circular and share a common
radius, upon perfect distortion correction. Figure~\ref{fig:sample_of_training_images__1}
shows a sample of training images that are gamma-corrected. The distorted
disks in Figs.~\ref{fig:sample_of_training_images__1}(a), (b), and
(e) lie on a distorted jittered Bravais lattices, simulating the effect
of Bragg diffraction from a crystal. The positions of the distorted
disks in Figs.~\ref{fig:sample_of_training_images__1}(c), (d), and
(f) were sampled uniformly in space. This was done for approximately
50\% of the images to prevent the DL model from predicting distortion
fields based on any features relating to reciprocal lattices, i.e.
features relating to the sample.

To describe the subspace of distortion fields from which we sampled
to generate our training and validation datasets, we introduce the
following quantities: $A_{\text{e}}=\sqrt{A_{r;2,0}^{2}+B_{r;1,0}^{2}}$,
$\phi_{\text{e}}=\text{atan2}\left(B_{r;1,0}/A_{r;2,0}\right)/2,$
$A_{\text{p}}=\sqrt{A_{r;1,1}^{2}+B_{r;0,1}^{2}}$, and $\phi_{\text{p}}=\text{atan2}\left(B_{r;0,1}/A_{r;1,1}\right)$.
Table~\ref{tab:distortion_parameter_distributions__1} lists the
distributions from which the distortion parameters $A_{r;0,2}$, $A_{t;0,2}$,
$A_{\text{e}}$, $\phi_{\text{e}}$, $A_{\text{p}}$, and $\phi_{\text{p}}$
are sampled. The sampling of $x_{c;D}$, and $y_{c;D}$ is more complicated,
though the distributions from which they are sampled are both approximately
equal to $\mathcal{U}_{\left[0.25,0.75\right)}$, i.e. the continuous
uniform distribution over the interval $\left[0.25,0.75\right)$. 

\begin{table}
\noindent \begin{centering}
\begin{tabular}{|c|c|}
\hline 
Distortion parameter & Distribution\tabularnewline
\hline 
$A_{r;0,2}$ & $\mathcal{U}_{\left[-0.3,1.5\right)}$\tabularnewline
\hline 
$A_{t;0,2}$ & $\mathcal{U}_{\left[-0.75,0.75\right)}$\tabularnewline
\hline 
$A_{\text{e}}$ & $\mathcal{U}_{\left[0,0.125\right)}$\tabularnewline
\hline 
$\phi_{\text{e}}$ & $\mathcal{U}_{\left[0,\pi\right)}$\tabularnewline
\hline 
$A_{\text{p}}$ & $\mathcal{U}_{\left[0,0.35\right)}$\tabularnewline
\hline 
$\phi_{\text{p}}$ & $\mathcal{U}_{\left[0,2\pi\right)}$\tabularnewline
\hline 
$x_{c;D}$ & $\sim\mathcal{U}_{\left[0.25,0.75\right)}$\tabularnewline
\hline 
$y_{c;D}$ & $\sim\mathcal{U}_{\left[0.25,0.75\right)}$\tabularnewline
\hline 
\end{tabular}
\par\end{centering}
\caption{The distributions from which the distortion parameters $A_{r;0,2}$,
$A_{t;0,2}$, $A_{\text{e}}$, $\phi_{\text{e}}$, $A_{\text{p}}$,
$\phi_{\text{p}}$, $x_{c;D}$, and $y_{c;D}$ are sampled, in generating
the training, validation, and testing datasets. $\mathcal{U}_{\left[a,b\right)}$
is the continuous uniform distribution over the interval $\left[a,b\right)$,
and $\sim\mathcal{U}_{\left[a,b\right)}$ is a distribution that is
approximately equal to $\mathcal{U}_{\left[a,b\right)}$. \label{tab:distortion_parameter_distributions__1}}
\end{table}

We generated $506880$ and $126720$ training and validation grayscaled
images respectively, each with dimensions of $512\times512$ in units
of pixels. Training and validation dataset generation was performed
using $\texttt{EMicroML}$, which makes heavy use of $\texttt{FakeCBED}$
\citep{Fakecbed_1} in this context. $\texttt{FakeCBED}$ is a Python
library that we have developed for generating images that depict the
essential geometric features of CBED patterns.

\subsection{Loss\label{subsec:loss__1}}

A popular choice for the single data instance loss for distortion
estimation problems is the end-point error (EPE) of the sampled distortion
field \citep{Liao_1,Li_1,Li_2}, where the EPE error of a sampled
vector field is defined as the Euclidean distance between the predicted
field vector and its corresponding ground truth, averaged over all
sampled positions $\left(u_{x},u_{y}\right)$. Such a loss is fine
for problems that consider generic mixtures of different types of
distortion as long as the distorted input images depict geometric
features that are perfectly correlated with all of the distortion
model parameters, and hence the distortion field described by said
parameters.

This is not the case for our particular problem as the generic mixtures
we consider also include the possibility of purely elliptical distortion.
Recall that, according to the assumptions made in Sec.~\ref{subsec:parameterizing_optical_distortion__1},
a CBED pattern containing CBED disks that are all perfectly circular
sharing a common radius, is a distortion-free CBED pattern. A distortion
field that implies an inverse mapping that transforms a distorted
CBED pattern into one that is distortion-free is a distortion field
that perfectly undistorts the distorted CBED pattern. For a CBED pattern
subject to purely elliptical distortion, there are an infinite number
of distortion fields that can perfectly undistort the CBED pattern,
each having the same elliptical distortion vector $\left(A_{r;2,0},B_{r;1,0}\right)$,
but a different distortion center. In other words, a distorted CBED
pattern obtained from a particular distortion field that is purely
elliptical will not depict geometric features that are perfectly correlated
with the distortion center of said field. 

\begin{figure}[h]
\noindent \begin{centering}
\includegraphics[scale=0.55]{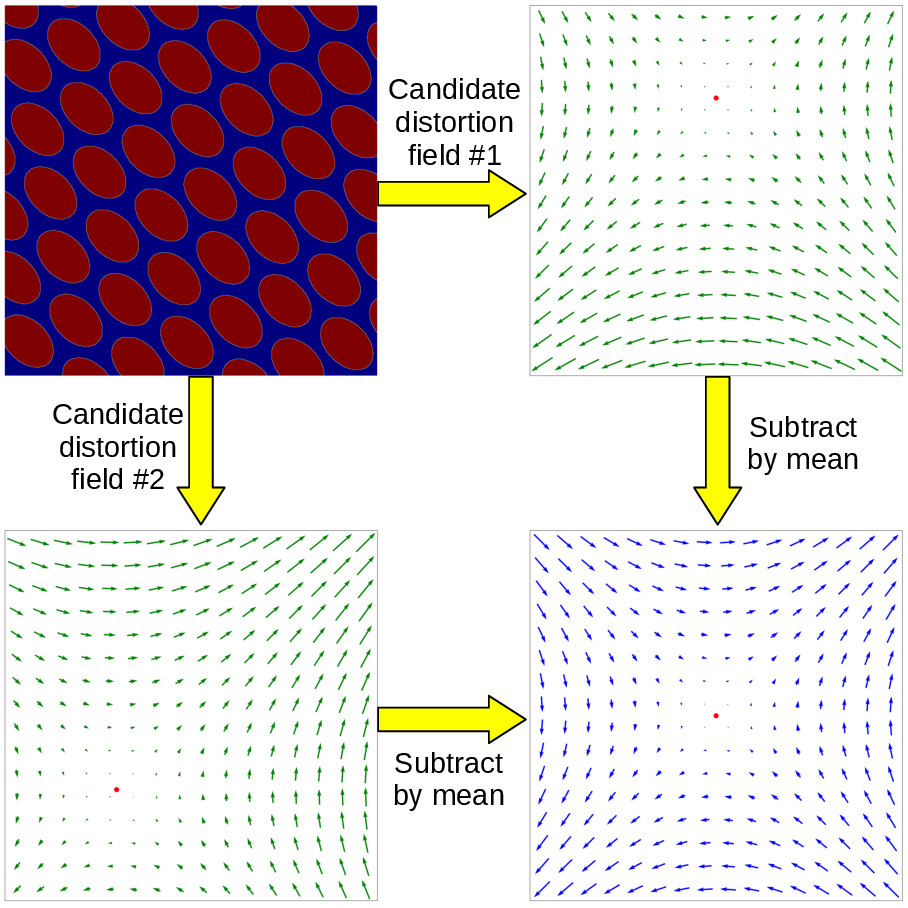}
\par\end{centering}
\caption{An illustration of the ambiguity of the distortion center for purely
elliptical distortion. The distortion fields in the bottom left and
top right corners are consistent with the distorted artificial CBED
pattern in the top left corner. However, subtracting these two distortion
fields by their respective means yield the same modified distortion
field, shown in the bottom right corner. The red dot in each distortion
field indicates its distortion center.\label{fig:ambiguity_of_center_of_elliptical_distortion__1}}
\end{figure}

Figure~\ref{fig:ambiguity_of_center_of_elliptical_distortion__1}
illustrates this ambiguity of the distortion center for purely elliptical
distortion. The difference between any two distortion fields consistent
with a CBED pattern subject to this type of distortion is a constant
vector field. Subtracting each of these distortion fields by their
respective means yields the same modified distortion field, which
incidentally is another distortion field consistent with the same
CBED pattern, with its distortion center coinciding with the center
of the distorted image. If any of the other types of distortion considered
in this paper are present in a given CBED pattern, then the ambiguity
vanishes and there is only one distortion field that is consistent
with the distorted CBED pattern. 

Consider an abstract undistorted CBED pattern of non-overlapping CBED
disks that share a common radius, and that outside the CBED disk supports
the intensity is zero, and inside each CBED disk support the intensity
is a common positive value. Next, consider an abstract distorted CBED
pattern obtained by distorting the abstract undistorted CBED pattern
according to a standard distortion field. Under certain circumstances,
as illustrated above, the abstract distorted CBED pattern will not
be perfectly correlated with the standard distortion field. However,
the abstract distorted CBED pattern should be perfectly correlated
with a vector field obtained by subtracting the original standard
distortion field by its mean. We refer to this vector field as the
``adjusted'' standard distortion field.

In light of the above remarks, we propose an alternative single data
instance loss $\mathcal{L}$ that can handle appropriately the special
case of purely elliptical distortion: the EPE of the sampled adjusted
standard distortion field. The mini-batch loss $\mathcal{L}_{B}$
is simply the single data instance loss averaged over all data instances
in the mini-batch. Each distortion field is sampled at the points
$\left(u_{x},u_{y}\right)\in\left\{ \left.\left(u_{x;m},u_{y;n}\right)\right|m,n\in\left\{ 0,\ldots,N-1\right\} \right\} $,
where $u_{x;m}=\left(m+1/2\right)/N$, $u_{y;n}=\left(n+1/2\right)/N$,
and $N$ is the number of pixels across the corresponding distorted
CBED pattern.

\subsection{Model training\label{subsec:model_training__1}}

Our DL model is trained via supervised learning using the mini-batch
stochastic gradient descent (SGD) optimization algorithm with the
weight decay (i.e. the L2 penalty) equal to $7.25\times10^{-4}$,
the momentum equal to $0.9$, and the mini-batch size equal to $64$.
The learning rate is updated after every processed mini-batch during
the training phase (i.e. not the validation phase). Over the first
$4$ epochs, the learning rate is increased linearly from $10^{-8}$
to $5\times10^{-3}$. Then, over the remaining $16$ epochs, the learning
rate is cosine annealed to $2\times10^{-5}$.

\begin{figure}

\noindent \begin{centering}
\includegraphics[scale=0.54]{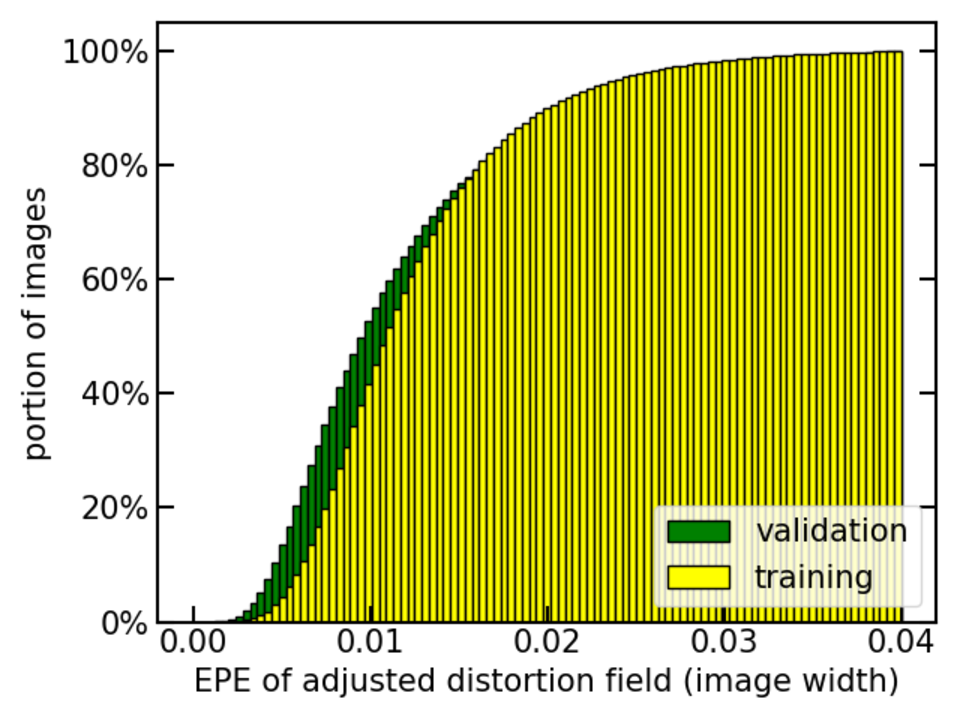}
\par\end{centering}
\caption{The cumulative distribution functions of the EPE of the adjusted distortion
field, in units of the image width, with the predictions made by our
trained DL model, for the training (yellow) and validation (green)
datasets.\label{fig:final_training_and_validation_epes__1}}

\end{figure}

Figure~\ref{fig:final_training_and_validation_epes__1} shows the
cumulative distribution functions (CDFs) of the EPE of the adjusted
distortion field for the training and validation datasets, after the
DL model has been trained. For both datasets, the DL model yields
EPEs of the adjusted distortion field less than $0.84\%$, $1.13\%$,
and $1.54\%$ of the image width for over $25\%$, $50\%$, and $75\%$
of the images respectively. Morever, we can see that no overfitting
has occurred during training.

\subsection{Testing data generation\label{subsec:testing_data_generation__1}}

\begin{figure}[h]
\noindent \begin{centering}
\includegraphics[scale=0.43]{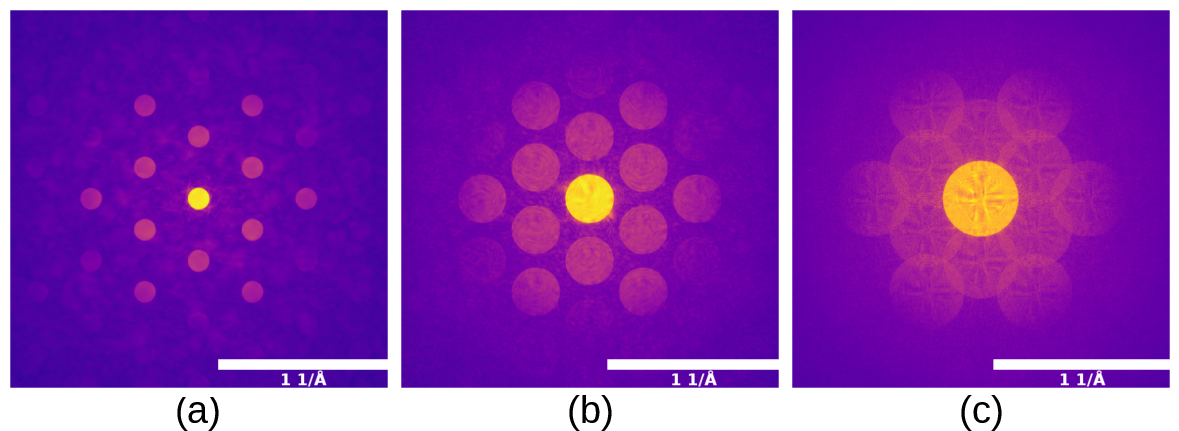}
\par\end{centering}
\caption{The three undistorted simulated CBED patterns used to generate the
testing datasets, gamma-corrected with a gamma value of $0.2$.\label{fig:undistorted_testing_images__1}}
\end{figure}

Given the way we generate our training dataset, it is natural to ask
whether our DL model, trained with such a dataset, can accurately
predict the distortion fields in images that more accurately depict
experimental CBED patterns, i.e. that are more representative of real
CBED patterns. To address this question, we generated test images
using multislice simulations. Specifically, we modelled a 5-layer
sub-specimen of $\ce{MoS2}$ on a $\unit[0.5]{nm}$ thick layer of
amorphous $\ce{C}$, that is illuminated by an electron beam operated
at $\unit[20]{keV}$. For all simulations, we included thermal effects,
Poisson noise, and spherical and chromatic aberrations. We considered
three semi-convergence angles, chosen to generate three target undistorted
CBED disk radii in units of the image width: $1/35$, $(1/35+1/10)/2$,
and $1/10$. For each semi-convergence angle, we generated a single
undistorted simulated CBED pattern, with dimensions of $512\times512$
in units of pixels. Figure~\ref{fig:undistorted_testing_images__1}
shows the three undistorted simulated CBED patterns. Note that for
Fig.~\ref{fig:undistorted_testing_images__1}(c), the CBED disks
are overlapping. 

For each undistorted simulated CBED pattern, we generated a testing
dataset containing $2880$ randomly distorted versions of said pattern,
where the distortion parameters were sampled in the same way as that
for the training and validation datasets. Moreover, random illumination
supports were imposed on the distorted simulated CBED patterns. Testing
dataset generation was performed also using $\texttt{EMicroML}$,
which makes heavy use of $\texttt{Prismatique}$ \citep{Prismatique_1}
in this context. $\texttt{Prismatique}$ is a Python library that
we have developed, that is a wrapper to the multislice Python library
$\texttt{PyPrismatic}$ \citep{DaCosta_1}. For convenience, we refer
to the testing datasets containing the randomly distorted versions
of the undistorted simulated CBED patterns shown in Figs.~(a), (b),
and (c) as testing datasets A, B, and C, respectively.

\subsection{Radial gradient maximization\label{subsec:radial_gradient_maximization__1}}

The performance test results of our DL approach are benchmarked against
those obtained using the RGM approach to distortion estimation, specifically
the single-disk variant \citep{Muller_1,Mahr_1}. Strictly speaking,
the RGM technique is used to estimate the displaced centers of the
CBED disks in the simulated CBED patterns generated in Sec.~\ref{subsec:testing_data_generation__1},
namely the CBED disks corresponding to the direct beam, and the Bragg
reflections that are nearest and next-nearest neighbours to the direct
beam (13 disks in total). For each simulated CBED pattern, the distortion
field is estimated as follows:
\begin{enumerate}
\item Estimate the displaced centers $\left\{ \left(\tilde{q}_{x:\text{C};i},\tilde{q}_{y;\text{C};i}\right)\right\} _{i=0}^{12}$
of the aforementioned $13$ CBED disks, where $\left(\tilde{q}_{x;\text{C};i},\tilde{q}_{y;\text{C};i}\right)$
is the estimated displaced center of the $i^{\text{th}}$ CBED disk;
\item Calculate Euclidean distances $\left\{ q_{\text{err};\text{C};i}\right\} _{i=0}^{12}$
between the estimated displaced centers $\left\{ \left(\tilde{q}_{x;\text{C};i},\tilde{q}_{y;\text{C};i}\right)\right\} _{i=0}^{12}$
and their respective ground truths $\left\{ \left(q_{x;\text{C};i},q_{y;\text{C};i}\right)\right\} _{i=0}^{12}$,
where $q_{\text{err};\text{C};i}=\sqrt{\sum_{\alpha\in\left\{ x,y\right\} }\left(\tilde{q}_{\alpha;\text{C};i}-q_{\alpha;\text{C};i}\right)^{2}}$;
\item Calculate an outlier registry $\left\{ \Theta_{\text{C};i}\right\} _{i=0}^{12}$,
where $\Theta_{i}$ is set to $1$ if $q_{\text{err};\text{C};i}$
differs from the mean of $\left\{ q_{\text{err};\text{C};i}\right\} _{i=0}^{12}$
by more than twice the standard deviation of $\left\{ q_{\text{err};\text{C};i}\right\} _{i=0}^{12}$,
else it is set to $0$;
\item Estimate the distortion parameters of the CBED pattern via non-linear
least squares, where the residuals $\left\{ \left.r_{\alpha;i}\right|\alpha\in\left\{ x,y\right\} \text{ and }i\in\left\{ 0,\ldots,12\right\} \right\} $
are calculated by $r_{\alpha;i}=\left[\tilde{q}_{\alpha;\text{C};i}-\tilde{D}_{\wasylozenge;\alpha}\left(u_{x}=u_{x;\text{C};i},u_{y}=u_{y;\text{C};i}\right)\right]\Theta_{\text{C};i}$,
with $\left(u_{x;\text{C};i},u_{y;\text{C};i}\right)$ being the center
of the $i^{\text{th}}$ CBED disk in the absence of distortion, and
$\tilde{D}_{\wasylozenge;\alpha}\left(u_{x},u_{y}\right)$ being the
$\alpha^{\text{th}}$ component of the distortion field $\left(\tilde{D}_{\wasylozenge;x}\left(u_{x},u_{y}\right),\tilde{D}_{\wasylozenge;y}\left(u_{x},u_{y}\right)\right)$
specified by the distortion parameters.
\end{enumerate}
The bencmarking results are presented in Sec.~\ref{subsec:multislice_simulation_tests__1}.

\subsection{Ptychography\label{subsec:ptychography__1}}

A computational approach to improving resolution that does not require
aberration correctors, or a high-pixel count detector is electron
ptychography \citep{Rodenburg_1}, which has been increasing in popularity
over the last two decades. Electron ptychography takes as input data
a 4D-STEM dataset, and constructs as output a model of the incident
electron beam and the electrostatic fields of the illuminated specimen,
i.e. the scattering object. Critically, the spacing between probe
positions must be sufficiently small such that there is an appreciable
amount of overlap between neighbouring illumination areas. This illumination
overlap imposes a mathematical constraint that is necessary in order
to construct a unique ptychographic model. As TEM and STEM images
only measure intensity, they discard the phase-shift information of
the exit electron waves that produce said images, and hence information
about the electrostatic fields in the specimen as well. By comparison,
a model of the scattering object, constructed via ptychography, estimates
this missing phase-shift information. Morever, from a model of the
scattering object, one can compute intensity images that quantify
the same information as that in TEM and STEM images.

The optical distortion present in the input 4D-STEM dataset must be
removed to provide the highest resolution reconstructions. In a previous
work of ours \citep{Blackburn_1}, an iterative distortion correction
technique was used that involved taking the Fourier transform (FT)
of a ptychographic reconstruction of a known material and fitting
this to a pincushion-distorted version of the expected (i.e. ideal
and undistorted) FT. Like the RGM approach, the iterative distortion
correction that we have previously used for ptychography requires
either a known calibration sample or good knowledge of the target
sample to be observed. Furthermore, this iterative approach assumed
only pincushion distortion, however there are other types of distortion
that may be present in the 4D-STEM data. 

As our first application of our DL framework, we correct the optical
distortion in an experimental 4D-STEM dataset used for electron ptychography
and show that this preprocessing of the data improves the quality
of ptychographic reconstructions, compared to those obtained using
the iterative distortion correction technique. The 4D-STEM dataset
was collected at $\unit[20]{keV}$ for a sample of $\ce{Au}$ islands
on a thin film of $\ce{MoS2}$. A subset of the CBED patterns in the
4D-STEM dataset were summed together to yield a pattern with disks
that has an enhanced signal-to-noise ratio compared to the individual
CBED patterns in the dataset. We assumed each CBED pattern in the
4D-STEM dataset was subject to the same distortion field, and estimated
said distortion field by passing the enhanced pattern through our
DL model. Distortion correction was applied to each CBED pattern in
the dataset via $\texttt{DistOptica}$, according to the distortion
field predicted by our DL model. The ptychographic reconstruction
was performed subsequently on the distortion-corrected dataset using
our own custom version of $\texttt{PtychoShelves}$ \citep{PtychoShelves_1}.
The results of this application are presented in Sec.~\ref{subsec:correcting_distortion_in_experimental_4D_STEM_datasets_for_ptychography__1}.

\subsection{Selected area electron diffraction\label{subsec:selected_area_electron_diffraction__1}}

SAED is one of the most common techniques for acquiring 2D electron
diffraction patterns. This technique can be used to determine the
structure, orientation, and defects of crystals. However, the accuracy
of the information that can be extracted from SAED is often limited
by the presence of optical distortion. Unlike CBED, the sample is
illuminated by a parallel or near-parallel electron beam in SAED,
which produces diffraction patterns containing diffraction spots rather
than diffraction disks. One can switch from a convergent to near-parallel
electron beam by adjusting the pre-specimen lenses only. Recall that
in Sec.~\ref{subsec:parameterizing_optical_distortion__1} we mentioned
that, broadly speaking, the distortion effects of pre-specimen lenses
are negligible compared to those of post-specimen lenses. This implies
that our DL framework can be adapted to SAED experiments as follows:
First, collect the target SAED data of the sample of interest; Second,
adjust the pre-specimen lenses only to form a convergent beam while
keeping the location of the direct beam fixed on the detector; Third,
collect CBED data and estimate the distortion field in the CBED data
using our DL framework; Finally, apply distortion correction to the
SAED data using the distortion field estimated in the previous step.

As our second application of our DL framework, we correct the distortion
in a SAED pattern of a single crystal $\ce{Au}$ specimen oriented
in the {[}100{]} direction, using the procedure described above. The
SAED experiment, like the corresponding CBED experiment, was performed
on the modified Hitachi SU9000 SEM described in Sec.~\ref{subsec:cbed__1},
operated at $\unit[20]{keV}$. The results of this application are
presented in Sec.~\ref{subsec:correcting_distortion_in_experimental_SAED_data__1}.

\section{Results and discussions\label{sec:Results_and_discussions__1}}

\subsection{Multislice simulation tests\label{subsec:multislice_simulation_tests__1}}

\begin{figure}[h]
\noindent \begin{centering}
\includegraphics[scale=0.43]{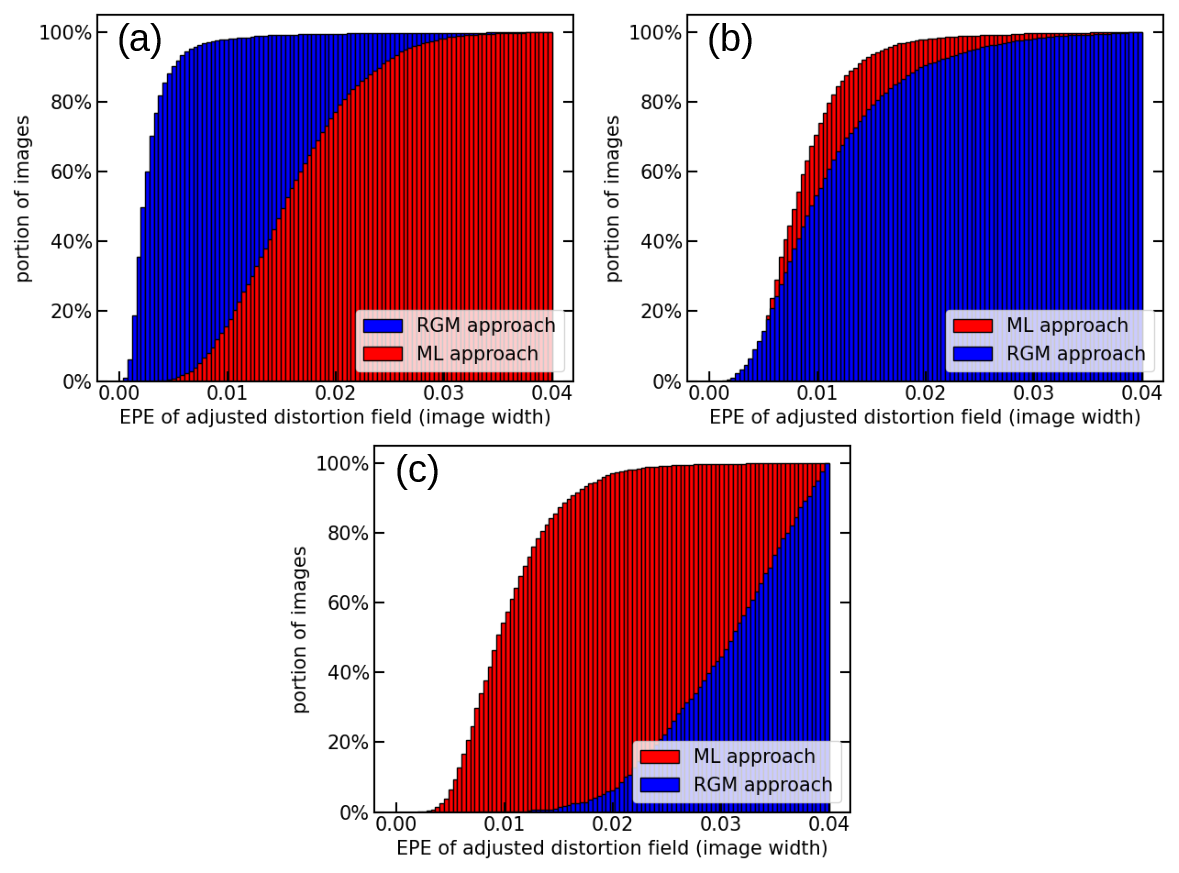}
\par\end{centering}
\caption{The cumulative distribution functions of the EPE of the adjusted distortion
field, in units of the image width, with the predictions made by our
DL model (red) and the RGM approach (blue), for: (a) testing dataset
A; (b) testing dataset B; and (c) testing dataset C. \label{fig:benchmarking_results__1}}
\end{figure}

Figure~\ref{fig:benchmarking_results__1} shows the cumulative distribution
functions (CDFs) of the EPE of the adjusted distortion field, with
the predictions made by our DL model, and the RGM approach, for the
three testing datasets that we generated using multislice simulations.
While the RGM approach outperforms our DL approach for the CBED patterns
with very small disks, our DL approach outperforms the RGM approach
for the CBED patterns with medium-sized disks, and large overlapping
disks. 

The fact that the RGM approach performs worse for the CBED patterns
with overlapping disks is expected, as the RGM technique relies on
the disks being well-separated. However, it is not obvious why the
performance of the RGM approach is appreciably better for the CBED
patterns with very small disks, compared to its performance for CBED
patterns with medium-sized disks. One possible explanation is that
the CBED disks corresponding to the Bragg reflections that are next-nearest
neighbours to the direct beam in Fig.~\ref{fig:undistorted_testing_images__1}(b)
appear to be less uniform compared to those in Fig.~\ref{fig:undistorted_testing_images__1}(a).
In general, the accuracy of the RGM technique should improve as the
CBED disks become more uniform. Upon further inspection, we have confirmed
that indeed the RGM technique does not predict the displaced centers
of these outer CBED disks as accurately in testing dataset B, as it
does for those in testing dataset A. A second possible explanation
is that the size of the search space of the displaced center of a
distorted CBED disk decreases as the radius of the CBED disk decreases,
hence the maximum possible error decreases as well. A third possible
explanation is that as the CBED disk radius decreases, smaller details
of the CBED disk deformities cannot be resolved in an image, thus
making it more difficult for our DL approach to extract geometric
features that are relevant to predicting distortion fields. According
to Fig.~\ref{fig:benchmarking_results__1}, our DL approach performs
best for CBED patterns of non-overlapping CBED disks of intermediate
size, with comparable performance for those of large overlapping CBED
disks. For testing datasets A, B, and C, the DL model yields EPEs
of the adjusted distortion field less than $1.96\%$, $1.06\%$, and
$1.27\%$ of the image width for over $75\%$ of the images respectively.

While there are cases where the RGM approach outperforms our DL approach,
it is important to stress that the latter has the advantage of not
requiring knowledge of the sample of interest. That being said, we
have demonstrated that there are conditions, common to many experiments,
under which our DL approach should outperform the RGM approach. The
benchmarking results in this section thus suggest that our DL approach
achieves a good compromise between convenience and accuracy.

\subsection{Correcting distortion in experimental 4D-STEM datasets for ptychography\label{subsec:correcting_distortion_in_experimental_4D_STEM_datasets_for_ptychography__1}}

\begin{figure}[h]
\noindent \begin{centering}
\includegraphics[scale=0.43]{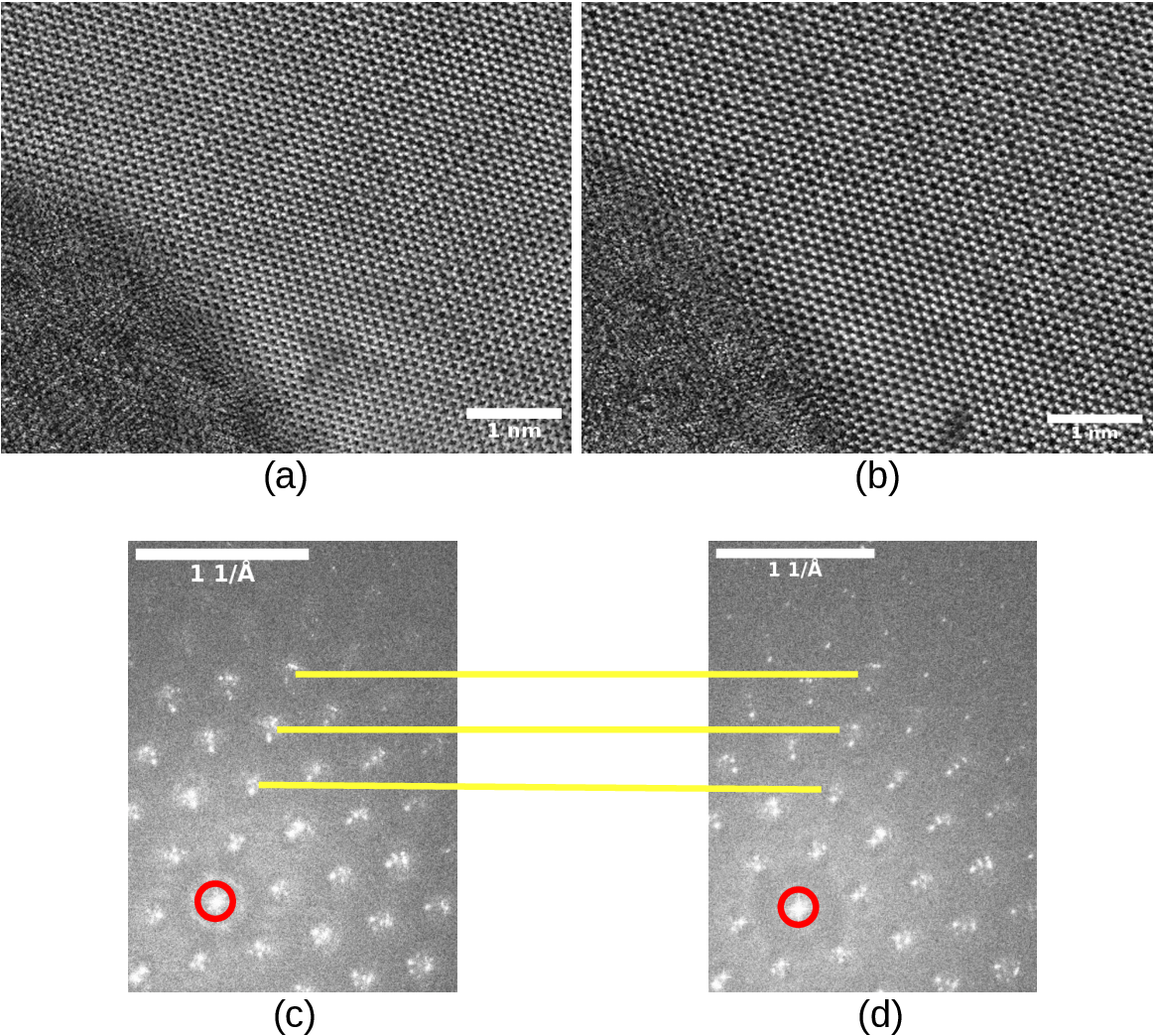}
\par\end{centering}
\caption{Ptychographic reconstructions using distortion-corrected experimental
4D-STEM datasets of a sample of $\ce{Au}$ on $\ce{MoS2}$: (a) The
amplitude of the exit wave resulting from the ptychographic reconstruction
using the 4D-STEM dataset obtained via the iterative distortion correction
method used in Ref.~\citep{Blackburn_1}; (b) Same as (a) except
the 4D-STEM dataset was obtained via our DL distortion correction
method; (c) The amplitude of the Fourier transform of (a); and (d)
The amplitude of the Fourier transform of (b). The yellow lines in
(c) and (d) guide the eye between related Fourier peaks. The red circles
mark the direct $\left(000\right)$ beams in the Fourier transforms.
\label{fig:ptychographic_reconstructions__1}}
\end{figure}

The amplitude of the exit wave resulting from the ptychographic reconstruction
using the experimental 4D-STEM dataset of the sample of $\ce{Au}$
on $\ce{MoS2}$, obtained via our DL distortion correction method,
is shown in Fig.~\ref{fig:ptychographic_reconstructions__1}(b).
For comparison, we show that which was obtained using the iterative
distortion correction method used in Ref.~\citep{Blackburn_1}, which
assumes that only pincushion distortion is present. Figures~\ref{fig:ptychographic_reconstructions__1}(c)
and (d) show the amplitudes of the Fourier transforms of (a) and (b)
respectively. The improved clarity in the amplitude of the exit wave
obtained using our DL distortion correction approach is apparent from
the sharper peaks exhibited in the corresponding Fourier transform.

\subsection{Correcting distortion in experimental SAED data\label{subsec:correcting_distortion_in_experimental_SAED_data__1}}

\begin{figure}[H]
\noindent \begin{centering}
\includegraphics[scale=0.43]{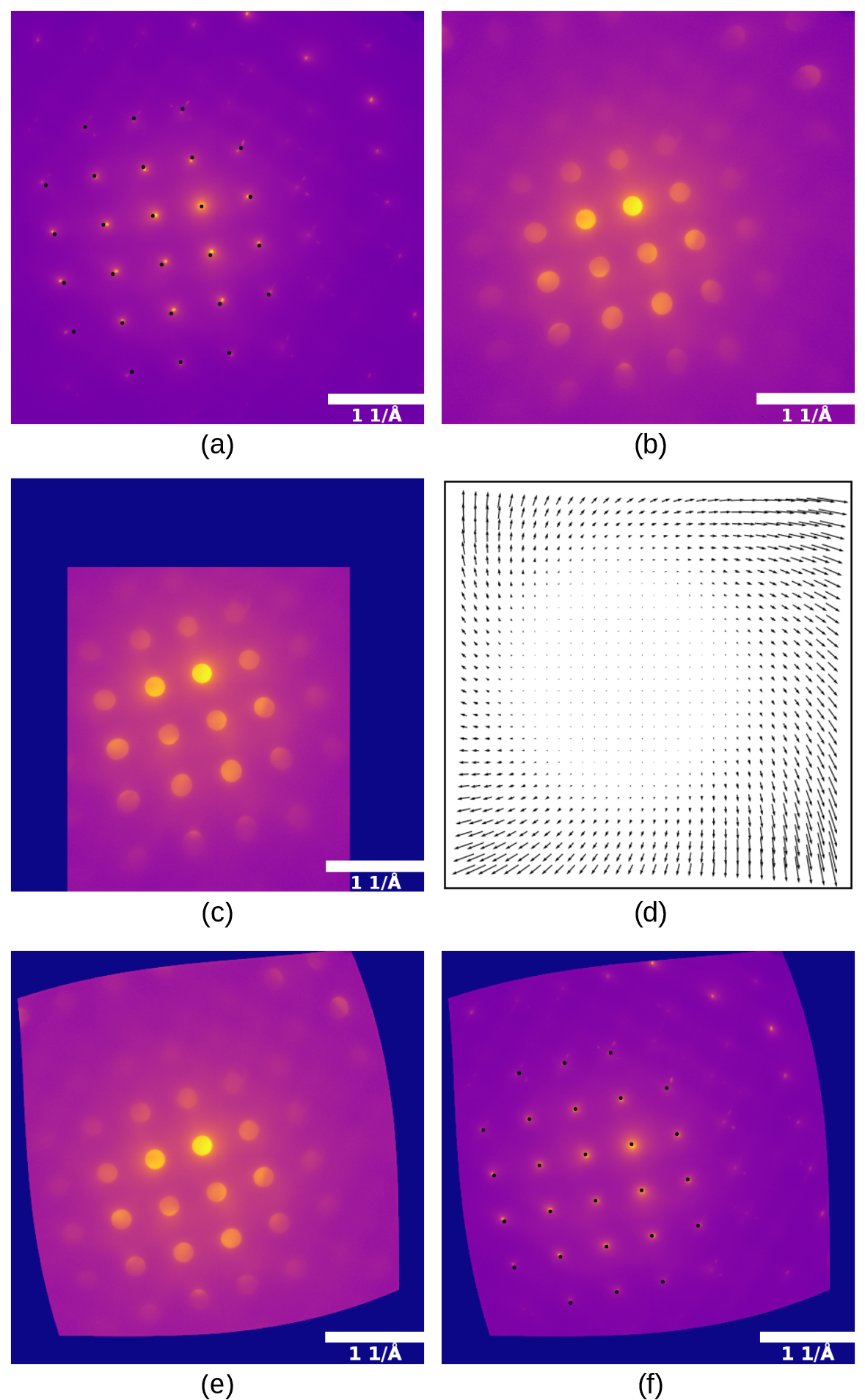}
\par\end{centering}
\caption{Correcting the distortion in an experimental SAED pattern of a single
crystal $\ce{Au}$ specimen oriented in the $\left[100\right]$ direction,
following the methodology described in Sec.~\ref{subsec:selected_area_electron_diffraction__1}:
(a) The as-recorded SAED pattern, prior to distortion correction;
(b) The as-recorded CBED pattern, prior to distortion correction;
(c) Same as (b) but with a rectangular mask frame; (d) The distortion
field of (c) predicted by our DL model; (e) The distortion-corrected
CBED pattern using the distortion field in (d); and (f) The distortion-corrected
SAED pattern using the distortion field in (d). In both (a) and (f),
the black dots form the best square lattice fits to the visible zero
order Laue zone reflections. All intensity patterns are gamma-corrected
with a gamma value of $0.2$. \label{fig:distortion_correction_of_saed_data__1}}
\end{figure}

Figure~\ref{fig:distortion_correction_of_saed_data__1} illustrates
the steps in our procedure for correcting the distortion in an experimental
SAED pattern of a single crystal $\ce{Au}$ specimen oriented in the
$\left[100\right]$ direction. The SAED pattern that we collected,
prior to distortion correction, is shown in Fig.~\ref{fig:distortion_correction_of_saed_data__1}(a).
After collecting the SAED pattern, we adjusted the pre-specimen lenses
to form a CBED pattern. This CBED pattern, prior to distortion correction,
is shown in Fig.~\ref{fig:distortion_correction_of_saed_data__1}(b).
Next, we apply a mask to block all but most of the zero-order Laue
zone (ZOLZ) reflections, which is shown in Fig.~\ref{subsec:selected_area_electron_diffraction__1}(c).
We found that masking can improve the performance of our DL model.
One possible explanation is that at low beam energies and small CBED
disk sizes, the Ewald sphere curvature can be quite pronounced and
the small-angle approximation may not hold across the entire angular
field of view of a given CBED pattern, both of which may affect the
validity of our assumption that the CBED pattern should depict only
near-perfect circular CBED disks of the same common radius, in the
absence of distortion. This should only be a concern at larger scattering
angles, which is why we did not mask most of the ZOLZ reflections
of (b). After masking, we used our DL model to predict the distortion
field of this CBED pattern, which is shown in Fig.~\ref{fig:distortion_correction_of_saed_data__1}(d).
Distortion correction was applied subsequently to the original CBED
pattern via $\texttt{DistOptica}$, according to the distortion field
predicted by our DL model. The distortion-corrected CBED pattern is
shown in Fig.~\ref{fig:distortion_correction_of_saed_data__1}(e).
Lastly, according to the same distortion field, we also applied distortion
correction to the original SAED pattern, the result of which is shown
in Fig.~\ref{fig:distortion_correction_of_saed_data__1}(f).

Given that our sample is single-crystal $\ce{Au}$ oriented in the
$\left[100\right]$ direction, used for calibration, we know that
the zero-order Laue zone (ZOLZ) reflections should lie approximately
on a square lattice. Again, the fact that they should lie only approximately
on a square lattice is due to the curvature of the Ewald sphere. However,
the deviation should be no more than $2$ pixels (or $0.0039$ in
units of the image width) for ZOLZ reflections. Therefore, to assess
the accuracy of the distortion correction, we fit squares lattices
to the most visible ZOLZ reflections, for each SAED pattern, which
are shown in Figs.~\ref{fig:distortion_correction_of_saed_data__1}(a)
and (f). We define the lattice fit error to be the square root of
the mean of the Euclidean distances squared between the ZOLZ reflections
and their corresponding points on the square lattice fit. 

The lattice fit shown in Fig.~\ref{fig:distortion_correction_of_saed_data__1}(f)
was obtained using the same DL model as that used in Secs.~\ref{subsec:multislice_simulation_tests__1}
and \ref{subsec:correcting_distortion_in_experimental_4D_STEM_datasets_for_ptychography__1}.
In units of the image width, the error of the fit for the as-recorded
SAED pattern, prior to distortion correction, was $0.0094$. For the
distortion-corrected SAED pattern, the error of the fit is $0.0046$,
which is more than a two-fold improvement. We also investigated how
sensitive the performance of our DL model was to the random number
generator (RNG) seed used during training. We trained $10$ DL models
in total, each with a different RNG seed, and found the mean of the
lattice fit error to be $0.0048$, with a standard deviation of $0.0017$.

The results in this section indicate that our assumption that optical
distortions originate predominantly from post-specimen lenses is reasonable.
Moreover, we have demonstrated in this section that our DL approach
to distortion correction is versatile enough to handle both CBED and
SAED data.

\section{Conclusions\label{sec:Conclusions__1}}

We have developed a DL framework for measuring and correcting optical
distortions in CBED patterns. We adopted a new method for training
data generation where we used basic mathematical functions to generate
images that depict the essential features of CBED patterns, rather
than use physics-based simulators. This approach enabled us to generate
large amounts of diverse training data fast, in a relatively straightforward
manner. However, with our approach, the training data set size, the
number of epochs, and the SGD weight decay hyperparameters, need to
be carefully chosen in order to avoid overfitting. It is important
to note that even if the distributions of the training and validation
losses are virtually identical at the end of training, there is still
the possibility of overfitting, as the training and validation datasets
contain artificial CBED patterns that capture only the essential geometric
features of experimental CBED patterns. The results of Sec.~\ref{sec:Results_and_discussions__1}
strongly indicate that we have successfully trained a DL model that
generalizes well to simulated CBED patterns that are generated using
physics-based simulators, as well as experimental CBED patterns.

It is worth emphasizing that the main feature of our DL approach to
distortion correction is that it does not require knowledge of the
sample of interest. By contrast, conventional distortion correction
approaches, e.g. the RGM approach, generally require either precise
knowledge of the sample being investigated, which is often not available,
or a calibration sample, which adds inconvenience to experiments.
While there are cases where the RGM approach outperforms our DL approach
in terms of accuracy, our benchmarking results of Sec.~\ref{subsec:multislice_simulation_tests__1}
show that there are conditions common to many experiments, e.g. ptychography
experiments, under which our DL approach outperforms the RGM approach.
We have also shown how our DL framework can be extended to handle
SAED patterns, thus demonstrating that not only is our approach to
the distortion correction of electron diffraction patterns accurate
and convenient, it is also very versatile. Our work is expected to
benefit high accuracy electron diffraction work, particularly at low-beam
energies, and improve the quality of electron ptychographic reconstructions,
which require accurate electron diffraction data.

\section*{CRediT authorship contribution statement\label{sec:credit_authorship_contribution_statement__1}}

\textbf{Matthew~R.~C.~Fitzpatrick}: Conceptualization, Methodology,
Software, Validation, Formal Analysis, Investigation, Data Curation,
Writing - Original Draft, Visualization. \textbf{Arthur~M.~Blackburn}:
Conceptualization, Software, Investigation, Writing - Review \& Editing,
Supervision, Project Administration, Funding Acquisition. \textbf{Cristina~Cordoba}:
Investigation, Writing - Review \& Editing.

\section*{Declaration of competing interest\label{sec:declaration_of_competing_interest__1}}

Hitachi High-Tech provided the SU9000 scanning electron microscope
used in this study, and supports a research chair for Arthur~M.~Blackburn.

\section*{Acknowledgements\label{sec:acknowledgements__1}}

We thank Kate Reidy and Frances Ross (MIT) for providing the $\ce{Au}$/$\ce{MoS2}$
sample. The extensive support given from Dectris AG (Switzerland)
and the provision of developmental Quadro cameras that expedited this
work is gratefully acknowledged. This work was part-funded by the
Natural Sciences and Engineering Research Council of Canada (NSERC),
from Collaborative Research and Development grant (CRDPJ 543431),
partnering with Hitachi High-Tech Canada. Part of the data processing
for ptychography was carried out using the cSAXS ptychography MATLAB
package developed by the Science IT and the coherent X-ray scattering
(CXS) groups, Paul Scherrer Institute, Switzerland. All code in this
work not pertaining to the acquisition of experimental data was performed
using the Advanced Research Computing (ARC) facilities of Digital
Research Alliance of Canada. Members of Hitachi High-Tech, Naka, Japan,
who developed and integrated the projection lens on the experimental
SU9000 instrument, are thanked and acknowledged, with special thanks
going to Kazutoshi Kaji, Satoshi Okada, and Toshi Agemura.

\section*{Data availability\label{sec:data_availability__1}}

The trained DL model used to obtain the results of Sec.~\ref{sec:Results_and_discussions__1},
along with all of the testing images, subsets of the training and
validation images, and the data presented in Figs.~\ref{fig:final_training_and_validation_epes__1}
and \ref{fig:benchmarking_results__1}, is freely available in the
Canadian Federated Research Data Repository at \href{https://doi.org/10.20383/103.01400}{https://doi.org/10.20383/103.01400}.

\section*{Code availability\label{sec:code_availability__1}}

All code in this work not pertaining to the acquisition of experimental
data nor ptychographic reconstructions is freely available in the
GitHub repository \href{https://github.com/mrfitzpa/emicroml}{https://github.com/mrfitzpa/emicroml}.

\appendix

\bibliographystyle{elsarticle-num}
\addcontentsline{toc}{section}{\refname}\bibliography{manuscript}

\end{document}